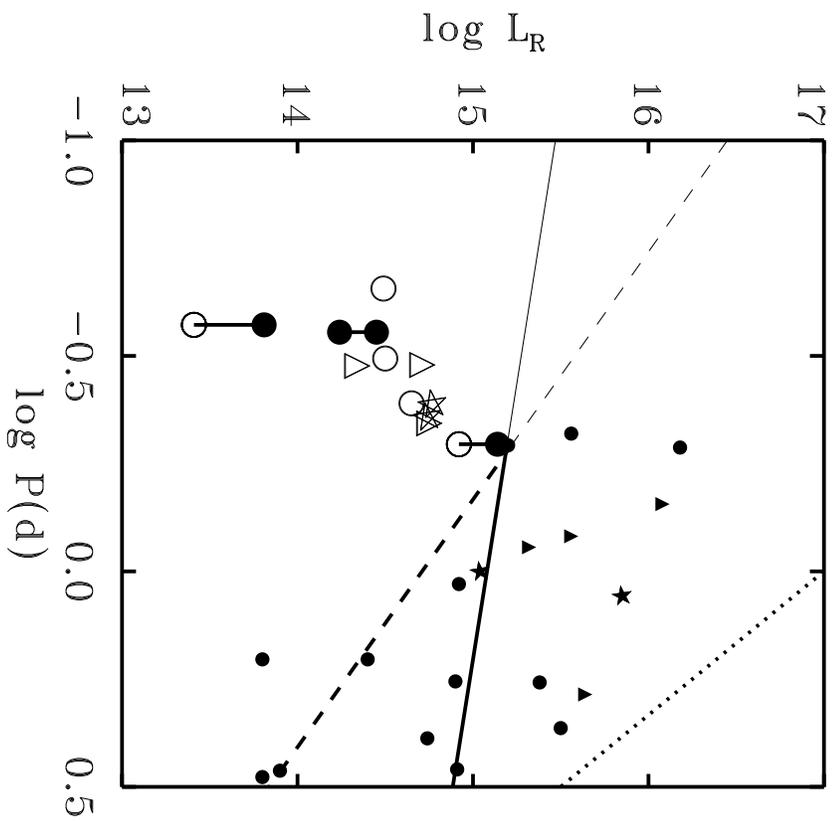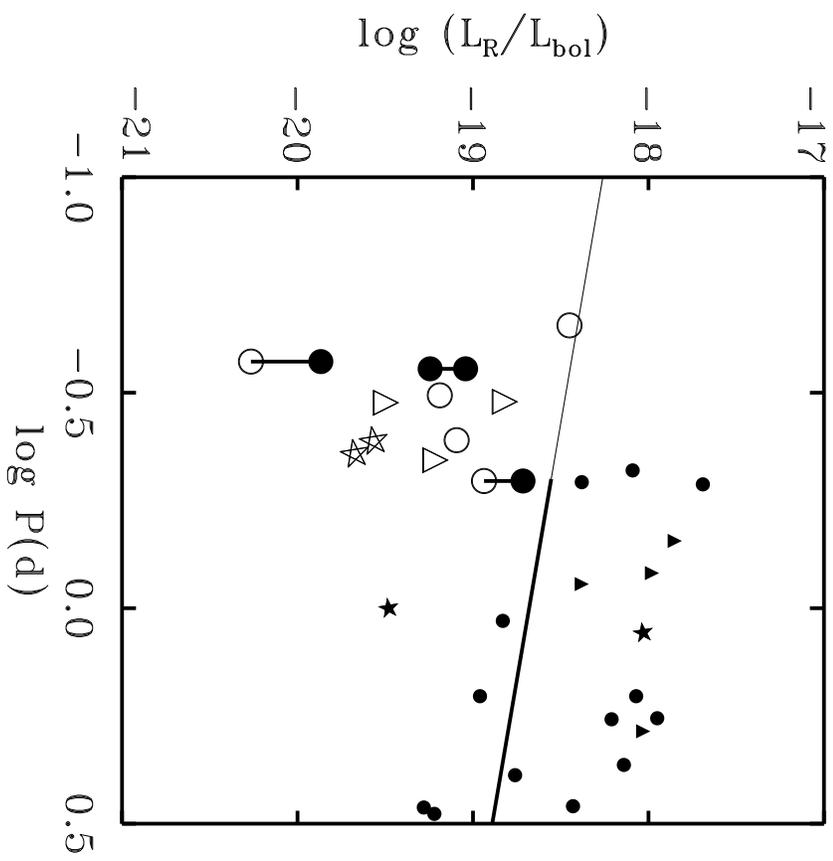

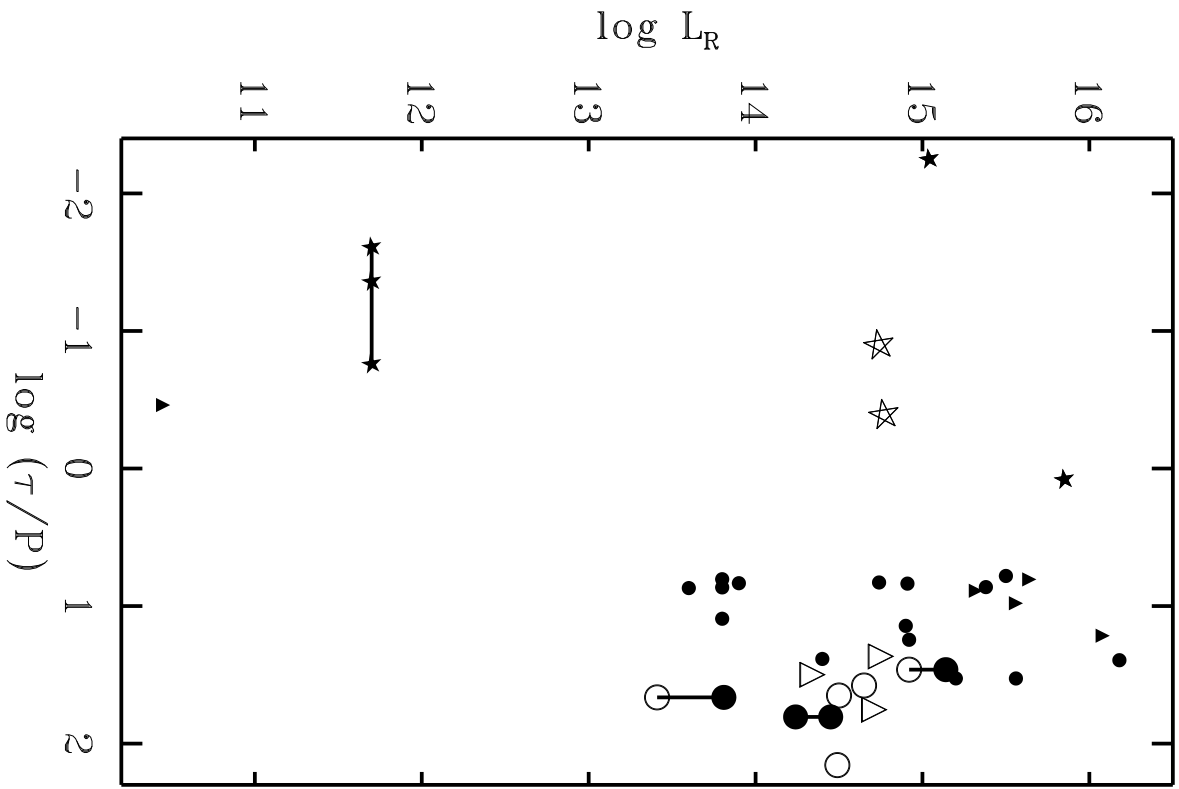
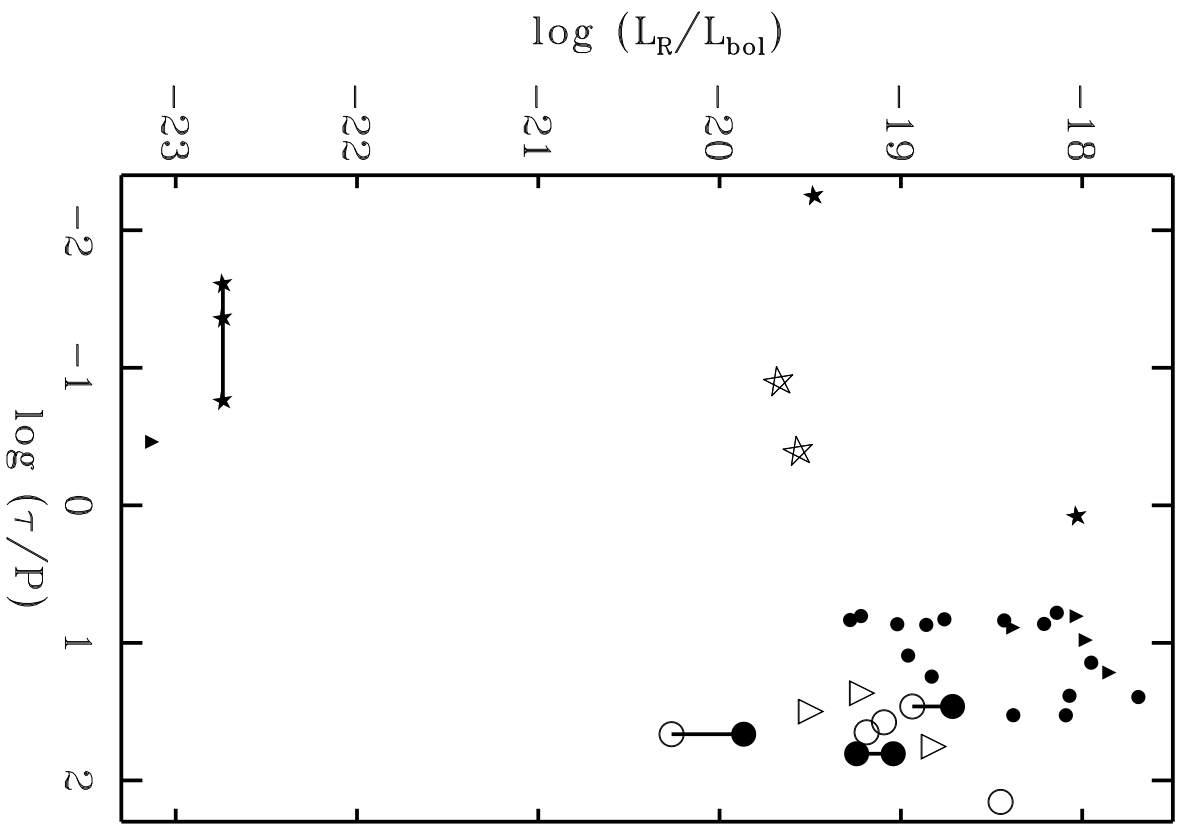

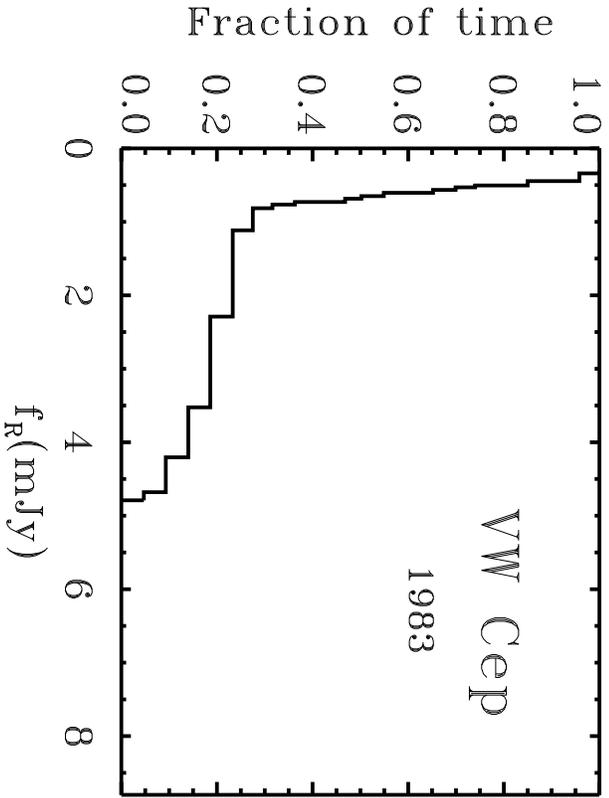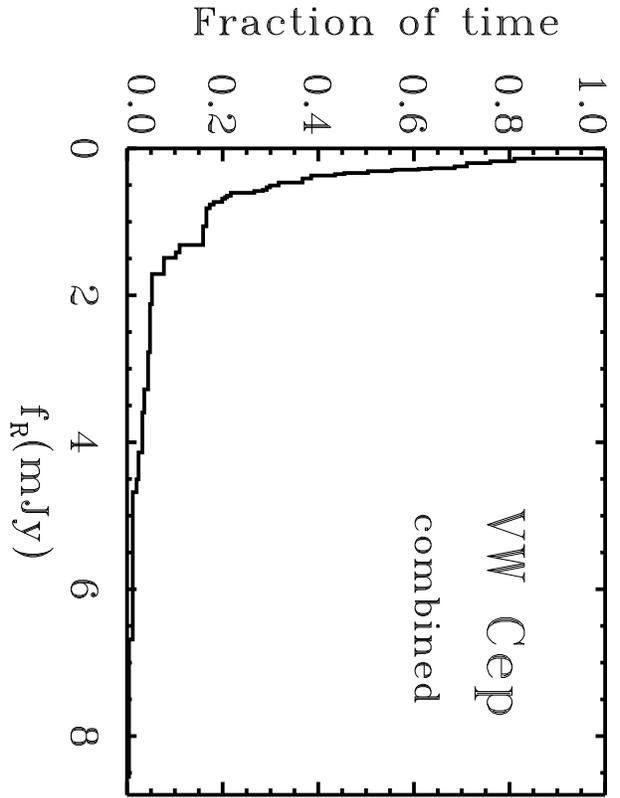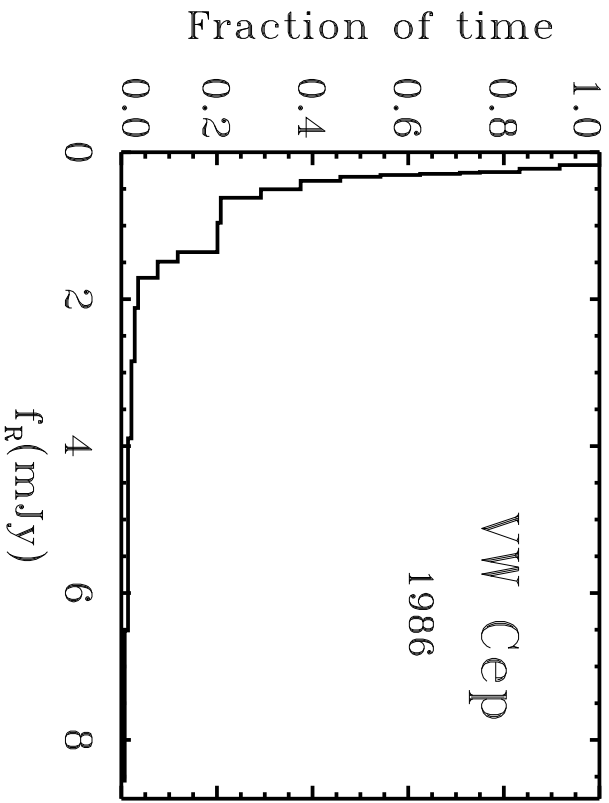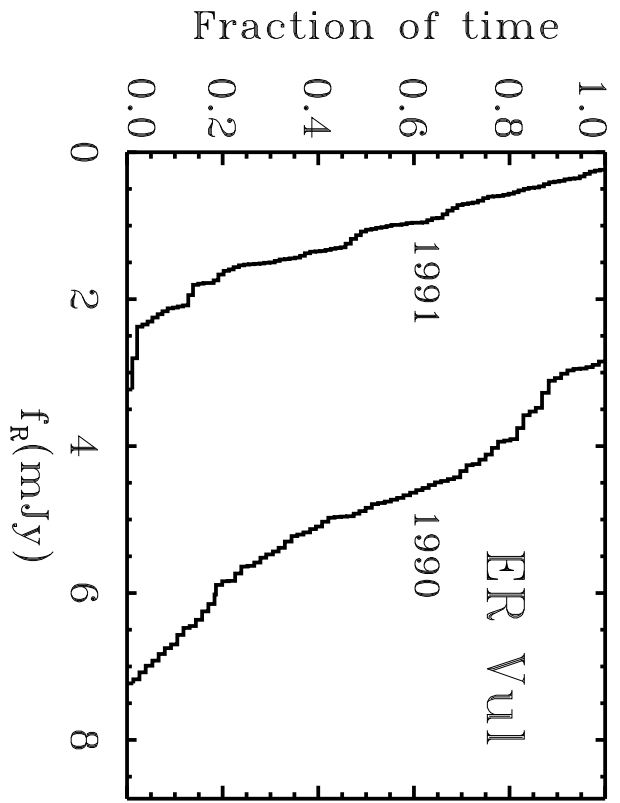

# Radio Survey of W UMa-type Systems

Slavek Rucinski[1]

*rucinski@astro.utoronto.ca*

81 Longbow Drive, Scarborough, Ontario M1W 2W6, Canada

February 17, 1995

## ABSTRACT

12 W UMa-type systems were observed at 3.6 cm with a goal to compare their radio luminosities with those of non-contact, rapidly rotating, late-type stars. Only 3 systems, all of spectral type K, have been detected. The detections and (low) upper limits for remaining systems confirm the strong under-luminosity of W UMa-type systems.

VW Cep, which was been detected several times before, has been analyzed in terms of the temporal cumulative flux distribution. Apparently, VW Cep spends most of the time in radio quiescence with flare-like events taking less than 20% of time.

## 1. Introduction

Contact binary stars of the W UMa-type have the shortest periods possible for binaries consisting of non-degenerate, Main Sequence stars. The orbital and rotational motions of the components are synchronized as they form single structures described by common equipotentials (Rucinski 1993). With spectral types ranging between A-F to early-K, they are expected to be the most active among solar-type stars. Yet, there is a growing evidence that their coronal activity is below that of short-period, synchronized, but detached binaries. The under-activity of W UMa-type systems in the coronal X-ray emission (first noticed by Cruddace & Dupree 1984) was explained by Vilhu & Walter (1987) by a spectral-type dependent saturation of activity at high rotation rates. Intensities of radio and X-ray emission are known to correlate for single stars and for components of detached binaries (Güdel & Benz 1993) so that scattered information on the under-activity of the W UMa systems in the radio (Hughes & McLean 1984, Drake et al. 1986, Gibson & Rucinski, 1985, unpublished, Rucinski & Seaquist 1988, Beasley et al. 1993) is consistent with the X-ray results.

Dwarf stars are faint and difficult targets for radio observations so that lack of detections could be simply due to insufficient sensitivity. Besides, most of the previous radio observations of the W UMa-type systems were done in a non-systematic way and included basically only bright systems on the sky, whereas optically fainter, late-type W UMa systems were not well represented. As a result, an important section of the relevant parameter space may have been left unexplored. This is because of the existence of the period – spectral-type correlation for the W UMa-type systems which results in *systematically more rapid rotation for fainter systems of later spectral types* (Rucinski 1993). This correlation produces a very large range of the Rossby-number parameter, $\tau/P$ (Rucinski 1985, Vilhu & Walter 1987), which relates the convective turn-over time, $\tau$, to the rotation period of the star, $P$, and is frequently used as a single quantity for correlations with activity levels (Noyes et al. 1984, Rucinski & VandenBerg 1990) since it is supposed to better represent the combined effect of the spectral type and rotation rate than the rotation period alone. However, is not obvious that it is a good measure in the radio as well. In fact, to our knowledge, no attempts have been made to relate radio luminosities of active stars to their Rossby number parameters. Note, that this parameter is sometimes called the "Rossby number" and sometimes the "inverse Rossby number".

---

[1]Affiliated with the David Dunlap Observatory, University of Toronto, P.O.Box 360, Richmond Hill, Ontario, Canada L4C 4Y6 and with the Department of Physics and Astronomy, York University, Toronto, Ontario, Canada.

The present mini-survey is an attempt to provide uniform data for a small but representative sample of contact binaries. The survey was conducted with the VLA[2] system, at 3.6 cm (X-band). At the VLA, this band combines high sensitivity with low confusion from bright extra-galactic sources; besides, since stellar radio spectra are known to be flat at high activity levels (eg. cases of FK Com, Rucinski (1991) or ER Vul, Rucinski (1992a)), the data could be directly comparable to previous results obtained mostly in the at 6 cm (C-band). In selecting the objects, an attempt was made to cover the whole range of spectral types observed for the W UMa systems, from early-F (AW UMa, V566 Oph) to early-K, with orbital sizes among the latter selected to include the compact (CC Com) as well as relatively large (OO Aql, AH Vir) systems.

Section 2 describes the VLA observations and the essential results are presented in Section 3. Section 4 contains a re-discussion of new and literature data for VW Cep, a system which was detected several times and whose variability might shed light on typical temporal behavior of W UMa systems. Section 5 gives a summary of the results.

## 2. VLA Observations

The observations were obtained during two VLA programs, AR 255 on April 24, 1992 and AR 318 on June 2, 1994. During both programs the same frequencies and bandwidths were used: these were two simultaneously observed 50 MHz-wide bands centered at 8415 and 8465 MHz. Calibrations to obtain linear polarization were not made. The antenna system was in the C configuration in April 1992 and in the B configuration in June 1994, resulting in the synthesized beam widths of typically $3''$ and $1''$, respectively. The observations were obtained by phase referencing to the VLA calibrators every 12 – 18 minutes, with total integration times of 24 to 48 minutes, as given in columns with headings "1992" and "1994" in Table 1. Because most of the selected W UMa systems are nearby objects, yet were expected to be faint in the radio, optical coordinates used for optical–radio cross-identifications were corrected for proper motions. Most of the proper motion data (with exceptions listed in Table 1) were taken from the SAO Catalogue.

---

[2]The Very Large Array is a facility of the National Radio Astronomy Observatory which is operated by Associated Universities, Inc., under cooperative agreement with the National Science Foundation.

The interferometric data were analyzed in a standard way using the AIPS software system[3], by obtaining maps over areas of $1024 \times 1024$ pixels (with pixel sizes $0''.25$ and $0''.1$) which were then "cleaned" using the AIPS task MX.

The first series of observations in 1992 lead to 3 detections for 7 targets observed. These 3 detected objects (44i Boo, VW Cep and OO Aql) were added to 4 remaining targets observed in 1994. The latter program gave only one detection, of VW Cep. Altogether, 12 W UMa systems were observed during both programs. Table 1 gives the positional and timing data whereas Table 2 gives the results of the observations. The values of the map $rms$ error per beam (Table 2) were used to estimate the upper limits for non-detections ($\approx 4 \times rms$) or errors of fluxes ($\approx 2 \times rms$). For the 4 detections, we give in Table 2 positions as well as shifts from the proper-motion-corrected optical positions of the objects. No attempt was made to detect circular polarization as all detections were at very low flux density levels.

## 3. Radio Activity of the W UMa Systems

### 3.1. Radio Luminosities

XY Leo has been eliminated from our analysis because its radio emission (Vilhu et al. 1988) might have been contaminated by that of its detached M-type binary companion (Barden 1987). From among the remaining 11 stars, we detected in 1992 only 44i Boo, OO Aql and VW Cep. These three were re-observed later, in 1994, but only VW Cep was seen again, in spite of comparable observing conditions and integration times (we discuss the case of VW Cep separately in Sec. 4).

There are basically two ways to estimate levels of radio emission for active stars: (1) one can evaluate radio luminosities, $L_R$, assuming distances to stars, or (2) one can relate the radio luminosities to the bolometric luminosities using the fact that the relative radio luminosity, $L_R/L_{bol}$, is equal to the ratio of the respective observed fluxes at Earth, $f_R/f_{bol}$. The second method gives a distance-independent measure of activity; in addition, it might be argued that it contains information on how much of the star's total energy becomes converted into its radio output. However, in reality, this fraction is always so small that $L_R$ might have little to do with the pool of the total available power radiated by the star, $L_{bol}$. Thus, if the latter quantity is really irrelevant here, by using $L_R/L_{bol}$, we might unnecessarily confuse the picture. Therefore, as a precaution, both measures, $L_R$ and $L_R/L_{bol} = f_R/f_{bol}$, have been used in our analysis.

---

[3]Astronomical Image Processing System (AIPS) is a data-reduction software package developed by National Radio Astronomy Observatory.

To evaluate the radio luminosities, $L_R$, the distances to our targets have been estimated using the recent absolute-magnitude calibration given in Rucinski (1994). This calibration is based on the orbital period, $P$ (measuring the size of the system) and on the $B - V$ color (measuring the effective temperature), and gives an estimate of the absolute visual magnitude, $M_V$. This latter quantity can be compared with the observed magnitude in the $V$-band to obtain the distance to the system, $d$, and $L_R = 4\pi d^2 f_R$. The input and derived data for the observed systems are listed in Table 3. The ordering of entries in this table is identical to that in Tables 1 and 2, where it follows the right-ascension and then the date of observations. The radio luminosity, $L_R$, is given in Table 3 in units erg s$^{-1}$Hz$^{-1}$. Consequently, the relative radio luminosity, $L_R/L_{bol}$, is measured in units of Hz$^{-1}$. The evaluation of $L_R/L_{bol} = f_R/f_{bol}$ was based on the observed visual magnitudes, $V$, and the bolometric corrections, $BC$ (from Popper 1980), using: $f_{bol} = 2.5 \times 10^{-5}\,\mathrm{dex}(-0.4(V + BC))$ erg s$^{-1}$. Note that for most objects we have only (relatively low) upper limits. The upper limits measured by Beasley et al. (1993) are some 5 times higher than ours so we decided not to use them here.

### 3.2. The period dependence

There exists no consensus at this time whether radio luminosities of active dwarfs depend in a simple way on their rotation periods. Stewart et al. (1988) argued that the rotation-period dependence of the *peak* luminosity for dwarfs of various (late) spectral types is well defined and relatively strong ($L_R^{peak} \propto P^{-3}$) while Drake et al. (1989) gave evidence that the period dependence of the *mean* luminosity is weak and shows much scatter ($L_R^{mean} \propto P^{-0.4}$). Since both measures of luminosity are correlated, this discordance is puzzling. More recently Güdel (1992) presented new data on *quiescent* radio emission of a more narrowly-defined group of active K-type dwarfs. This latter result is of obvious relevance for our detections only among K-type contact systems. Güdel again found a well defined and relatively strong dependence of the quiescent luminosity on the rotation period ($L_R^K \propto P^{-1.74}$). All these relations are based on stars with longer rotation periods than for W UMa systems, with the short period end defined by very few systems, such as AB Doradus ($P = 0.51$ day; Lim et al. 1992, Lim 1994) and ER Vulpeculae ($P = 0.70$ day; Rucinski 1992a). It is not obvious that these relations should apply to W UMa-type systems with periods as short as quarter of a day.

In Figure 1, we show our results for the W UMa-type systems (large symbols) together with individual detections for single stars and detached binaries from the surveys of Drake et al. (1989) and Güdel (1992) (small symbols), and with the linear relations derived by Stewart et al., Drake et al. and Güdel[4]. All detections are marked by closed symbols; our upper limits are marked by open symbols. The stars have been divided into three spectral-type groups, F, G and K, according to the $B-V$ color. The linear fits in Figure 1 may look as somewhat inconsistent with the data used to derive them but, (1) this figure contains combined samples of short period F–G–K-type stars from Drake et al. and of (different) K-type stars from Güdel and (2) the Drake et al. linear fits were derived for a sample covering a wider range of periods than shown in Figure 1. We did not make any distinction between the samples since it is not our intention to re-discuss the period dependence for short rotation periods. Rather, we would like to point out that whatever relation we use, the W UMa systems seem to be under-active for their short periods. The only detection in some agreement with the previous relations is that for OO Aql. This binary is *not* a typical W UMa system as it is larger and cooler than most of them, indicating that it has formed from a detached binary whose components were already somewhat evolved before coming into contact (Mochnacki 1981, 1985).

Not only do we not see any increase in radio activity for decreased periods in Figure 1, but the distribution of points for W UMa systems which we did detect might indicate a turn-over in the sense that *the W UMa systems become progressively less radio-active for shorter rotation periods*. This assertion is based on a very small range of periods so it cannot be taken as proven; possibly, there is no relation there at all. We note, however, that similar turn-over behavior was noted before for chromospheric and transition-region fluxes by Vilhu & Rucinski (1983) and by Rutten & Schrijver (1987).

The very recent detection of a radio-active star, 47 Cas (Güdel et al. 1995) has added a new important data point: This early F-type star follows the Drake et al. and Güdel fits for $\log L_R$ but its relative radio luminosity, $\log(L_R/L_{bol})$, is similar to those observed for the W UMa-type systems. This is visible in Figure 1 where 47 Cas has been marked by a filled asterisk at $P = 1$ day. We have only upper limits for the two F-type W UMa systems but they are very close to the level observed for 47 Cas.

### 3.3. The Rossby-number dependence

In Figure 2 the same data points as in Figure 1 are shown versus the Rossby number parameter, $\tau/P$. The numbers used here were determined using the formula for the convective turn-over times $\tau$ given by Rucinski & VandenBerg (1990). Since this formula applies only to $0.4 < B-V < 0.95$, we used a simple linear extrapolation to $\log \tau(1.4) = 1.59$. Application of this formula is particularly disputable for 47 Cas with $B - V = 0.31$; for this star we obtain $\tau = 8$ minutes only.

---

[4]Only Güdel (1992) published the numerical data for the regression line $\log L_R^K = 14.71 - 1.74 \log P$ (the rotation period $P$ is in days). The approximate shape of the Stewart et al. (1988) dependence has been estimated from Fig. 1 in their paper: $\log L_R^{peak} = 17 - 3 \log P$. Dr. Stephen Drake kindly provided the data for relations derived by Drake et al. (1989) and shown graphically in their paper: $\log L_R^{mean} = 15.08 - 0.39 \log P$ and $\log(L_R^{mean}/L_{bol}) = -18.68 - 0.42 \log P$.

Figure 2 is somewhat expanded relative to Figure 1 since we wanted to accommodate the enormous range in the Rossby numbers covered by the W UMa systems. For contact systems of F-type, the Rossby parameter is as small as that for the Sun, whereas for short-period K-type systems the range in this number extends beyond anything observed for non-contact systems.

In addition to the Sun plotted for $P = 26$ days, $B - V = 0.63$ and $\log L_R = 10.45$, we have added also Procyon ($B - V = 0.42$) and 47 Cas as the only two currently-known radio-detected F-type dwarfs. Procyon was detected by Drake et al. (1993) with $\log L_R = 11.7$. To estimate the Rossby parameter for Procyon, we used an approach which was in fact criticized by Soderblom (1985), but which should give at least a ball-park estimate of its rotation period: Using $V \sin i = 2.8$ km s$^{-1}$ from Gray (1981), we obtained the equatorial velocity, 5.3 km s$^{-1}$, assuming that the inclination is the same ($i = 32°$) as that for the orbit with the white-dwarf companion (Irwin et al. 1992). With a linear mass–radius relation, and $M = 1.75\,M_\odot$, an estimate of the period is 17 days. Recently, Hale (1994), determined $V \sin i = 5.0 \pm 1.0$ km s$^{-1}$ which would result in the rotation period of 9 days. Finally, Campbell and Garrison (1985) derived a period as short as 2.4 day. The values of $\tau/P$ that we obtained are then 0.024, 0.044 and 0.17. We note that Simon & Drake (1989), using similar guesses obtained 0.12.

The difficulties in estimating the convective turn-over times for F-type stars with shallow convective envelopes are even more acute in the case of 47 Cas which has an earlier spectral type than Procyon. We assumed that its period of rotation is one day, as suggested by Güdel et al. (1995) resulting in $\log(\tau/P) = -2.25$.

Figure 2 is not easy to interpret. On one hand our detections for the K-type W UMa systems still indicate some under-luminosity relative to normal stars, but the difference seems to be less pronounced than on the period plots in Figure 1. Since we do not know the functional dependences in Figure 2, we cannot really meaningfully compare the activity levels for the contact and non-contact samples. For $\log(\tau/P) < 0$, where we have only F-type systems (except the Sun), the situation is even less clear: The activity levels of Procyon and of 47 Cas are very different and we have no idea what are the "standard" levels to use for comparison with the high upper limits for the two F-type W UMa systems. In fact, it might be that the Rossby-parameter picture is not appropriate in attempts to understand activity of very rapidly-rotating stars, and over such a wide range of spectral types, where $\tau$ ranges between fraction of an hour and weeks. Although Figure 2 might contain important hints, the enormous range in the Rossby parameter and strong selection effects against discovery of faint sources, which are still present in all surveys of stellar radio activity, make interpretation of the functional dependence of the radio emission of contact binaries on Rossby number premature at this time.

## 4. Quiescent Emission of W UMa Systems: The Case of VW Cep

VW Cep was detected at the VLA each time it was observed, most probably because it is one of the nearest W UMa systems. As such, it offers a good test case to study the temporal flux distribution of the radio emission from a contact system. We use here all published data for VW Cep, obtained in 4 programs: Hughes & McLean (1984), Rucinski & Seaquist (1988), Vilhu et al. (1988) and the present. The published data are not exactly in a format which one would like to have when analyzing the temporal distributions of the flux. What is available can be summarized as: "during an integration of so many minutes, the average flux of VW Cep was so many milli-Janskys". However, this is the only type of information on temporal behavior of VW Cep in hand at this moment. However, it does contain the essential time information as observers normally either divide integrations into smaller sections when the flux is high and combine many integrations when the flux is low. A somewhat extreme case was that of Rucinski & Seaquist (1988) who observed the system for 24 hours in 20 cm, 6 cm and 2 cm spectral bands to study its time variability and did not detect the source in any of individual 12 minute integrations; they combined all integrations and detected the source at 6 cm at a very faint level of 0.11 mJy. This low flux corresponds to a low radio luminosity $\log L_R = 13.94$ ($L_R$ in $erg\ s^{-1}\ Hz^{-1}$) and might represent the real quiescent activity level of VW Cep.

The available data for VW Cep are listed in the first two columns of Table 4. There are 39 separate integrations given there, both at 3.6 cm and at 6 cm, ordered by the level of the flux. In combining the data, we have assumed that the spectrum is flat between these two wavelength bands. We can obtain information about typical flux levels for VW Cep by adding up the time when the star was observed to be brighter than each level (third and fourth columns of Table 4). This is shown graphically in the upper left panel of Figure 3 which gives the temporal cumulative flux distribution for the total integration time of 1209 minutes. During this time, the star was mostly faint and for only about 20% of the time was it brighter than $\simeq 0.5$ mJy. The mean flux level obtained from data in Table 4 is 0.62 mJy whereas the median is 0.32 mJy. These levels correspond to the radio luminosities $\log L_R$ (in erg s$^{-1}$ Hz$^{-1}$) of 14.70 and 14.41, respectively.

The presence of a break at about 0.5 mJy in the temporal flux distribution in Figure 3 suggests the possibility of a change in the radiation mechanism, from a quiescent or slowly varying component, to a rapidly changing, flare-like component. The break seems to be a permanent feature and is actually also visible to some degree in separate distributions formed from the individual datasets of Hughes & McLean (1984) and Vilhu et al. (1988) (see lower panels in Figure 3). Thus, we might suspect that the pattern is not variable in time (as with, say, the dynamo cycle phase) and is characteristic for the star itself.

It would be interesting to compare VW Cep with other very active stars, but this is not easy. Contrary to several RS CVn-type stars which were observed over long periods of time, few rapidly-rotating, late-type dwarfs have been observed with comparably long total integrations as that used for VW Cep. The only frequently observed early-K dwarf with which we can compare

VW Cep is AB Dor, a young star with rotation period of 0.51 day. As was shown by Lim et al. 1992, 1994) this star occasionally shows rotationally-modulated flaring emission, but spends most of its time, probably also roughly 80% in "quiescence", so that in terms of its temporal behavior it does not differ markedly from VW Cep (Lim 1994).

The data which we have in hand for a pair of rapidly-rotating G-type dwarfs, the close, synchronized binary ER Vul with $P = 0.70$ day (Rucinski 1992a), give an entirely different picture than for VW Cep. This is shown in the upper right panel of Figure 3. The system was at different levels of activity in 1990 and 1991, but each time the flux distribution did not indicate any clear separation into a quiescent and flare components. It is quite difficult to say at this point whether the difference between VW Cep and ER Vul is quite accidental or whether it indicates a deeper underlying disparity between magnetic processes in both binary systems. Unfortunately, the heterogeneous nature of the VW Cep data which were obtained over several years precludes any chances of rigorous statistical comparison of its temporal distributions with those for ER Vul which were based for each season on two consecutive days of continuous observations. At this point, we can only signal this potentially important difference.

## 5. Conclusions

A survey of 12 W UMa systems gave only 3 detections, with VW Cep detected twice. It is significant that all three stars are of K spectral type. The detections and upper limits for the remaining systems confirm the strong under-activity of contact binaries when compared with single, rapidly-rotating stars and components of short-period, synchronized binaries. This result fully confirms the conclusions of Beasley et al. (1993); the under-activity is now well established and is not an artifact of a small, unrepresentative sample or insufficient sensitivity. Although reasons for this effect could, possibly, be related to decreased differential rotation in contact stars, as suggested by Beasley et al., there could be as well other reasons for under-activity. One such possibility are shallower convective zones for contact components relative to normal stars (Rucinski 1992b). Another possibility is the influence of enormous energy transfer between components in contact binaries. Such transfer is apparently not affecting photospheric layers of contact binaries but must be present underneath, as there is no other way to explain the equality of effective temperatures for typically very dissimilar components (Rucinski 1993).

The well observed case of VW Cep gives us a chance to look at the temporal characteristics of its radio activity. The star spends most of the time in very low quiescence and only for about 20% of time shows flare-like brightenings. Possibly, such temporal characteristics – which are different than for a close, detached, synchronized binary ER Vul – hold a clue to the radio under-activity of the W UMa-type systems.

Thanks are due to Jean-Pierre Caillault for sending the numerical data for results on VW Cep

– 9 –

published in graphical form by Vilhu et al. (1988), to Stephen Drake for sending the unpublished cofficients of the period – luminosity relations and to David Gray for useful references. Thanks are also due to Ernie Seaquist and especially Stephen Drake for reading the manuscript and for useful comments and suggestions.

Special thanks are due to my wife, Anna, for financial support of my research.

The research grant from the Natural Sciences and Engineering Council of Canada is acknowledged with gratitude.

Captions for figures:

1. The radio luminosities, $L_R$ (in erg s$^{-1}$Hz$^{-1}$), and relative radio luminosities, $L_R/L_{bol}$ (in Hz$^{-1}$), are shown in relation to the orbital period of the system, $P$ (in days) for the W UMa-type systems studied here (large symbols) and for active dwarfs in surveys of Drake et al. (1989) and Güdel (1992) (small symbols). The detections are marked by filled symbols, whereas open symbols give upper limits. The spectral types are coded by shapes of symbols, according to ranges of the $B - V$ color: asterisks for F-type ($B - V < 0.5$), triangles for G-type ($0.5 < B - V < 0.7$) and circles for K-type ($B - V > 0.7$).

The slanted lines give the following rotation-period dependences: the *average* $L_R^{mean}$ and $L_R^{mean}/L_{bol}$ for a mixture of active stars by Drake et al. (1989) (the solid lines), the *quiescent* $L_R^K$ for rapidly-rotating, single, detached-binary K-type dwarfs by Güdel (1992) (the dashed line), and the *peak* $L_R^{peak}$ for active stars by Stewart et al. (1988). Extentions of the relations below $P = 0.5$ day marked by thin lines to indicate that they are extrapolations beyond the period range that they were based on.

Note that the evolved, long-period W UMa-type system, OO Aql ($P = 0.507$ day), on one occasion was as bright as predicted by the Drake et al. or Güdel relations, but for most of the time the W UMa systems are faint in the radio. The relatively high upper limit on $L_R/L_{bol}$ for CC Com (the shortest period) is due to the low bolometric luminosity of this very compact and cool system. Note also that the newly detected 47 Cas (Güdel et al. 1995), which shows high activity, for an early F-type star (asterisk at the period of one day), has $L_R$ at the level observed for most of the non-contact stars but its relative radio luminosity is comparable to levels observed for the W UMa-type systems.

2. The same quantities as shown in Figure 1, with the same symbols, but using the Rossby parameter, $\tau/P$, as the independent variable. This figure is somewhat expanded relative to Figure 1 to accomodate stars showing low radio activity, the Sun (a small triangle at the lowest edge) and Procyon (asterisks at the lower edge marking the Rossby parameter for periods 17, 9 and 2.4 day), as well as the active F-type star 47 Cas (asterisk at the left edge).

3. The temporal cumulative flux distribution for VW Cep for the combined dataset (upper left panel) is based on several separate observations resulting in total observing time of 20 hours. There are indications that this shape of the distribution is typical for the star and does not depend on the epoch of observations as distributions formed for subsets by Hughes & Mclean (1984) in 1983 (total $3^h24^m$) and by Vilhu et al. (1988) in 1986 (total $12^h$) have very similar shapes (lower two panels). The distributions for the detached, close, synchronized binary ER Vul consisting of two identical G-type dwarfs (Rucinski 1992a) have very different shapes (upper right panel). The ER Vul distributions were obtained on the basis of two programs, each time performed on two consecutive days, so that a strict comparison with the heterogeneous material for VW Cep is not possible.

Table 1. VLA Observations of W UMa Systems

| System | SAO | $\alpha_{1950}$ | s/year | $\delta_{1950}$ | ″/year | 1992[a] | 1994[b] | UT | UT |
|---|---|---|---|---|---|---|---|---|---|
| W UMa | 27364 | 9 40 15.4 | +0.0030 | 56 10 56.3 | −0.029 | | 48 | | 23:06–00:01 |
| XY Leo[c] | | 9 58 56.0 | | 17 39 02.0 | | | 48 | | 00:08–01:03 |
| AW UMa | 62579 | 11 27 25.6 | −0.0077 | 30 14 35.2 | −0.203 | | 48 | | 02:11–03:06 |
| CC Com[d] | | 12 09 33.8 | −0.0092 | 22 48 39.0 | +0.010 | | 48 | | 03:14–04:08 |
| AH Vir | 100003 | 12 11 47.9 | +0.0015 | 12 05 55.3 | −0.111 | | 48 | | 01:10–02:05 |
| 44i Boo | 45357 | 15 02 08.3 | −0.0409 | 47 50 53.3 | +0.032 | 24 | 31 | 11:08–11:36 | 04:15–04:50 |
| V502 Oph | 121784 | 16 38 47.8 | −0.0012 | 00 36 08.5 | +0.020 | 24 | | 11:46–12:14 | |
| V566 Oph | 122946 | 17 54 24.3 | +0.0046 | 04 59 30.8 | +0.080 | 24 | | 12:24–12:52 | |
| OO Aql | 125084 | 19 45 48.4 | +0.0044 | 09 11 01.5 | −0.003 | 24 | 31 | 13:02–13:30 | 05:42–06:17 |
| VW Cep | 9828 | 20 38 03.0 | +0.0899 | 75 24 58.4 | +0.560 | 24 | 31 | 13:40–14:08 | 04:56–05:31 |
| SW Lac | 72820 | 22 51 22.5 | +0.0058 | 37 40 18.9 | +0.005 | 24 | | 14:18–14:46 | |
| AB And | 73069 | 23 09 08.6 | +0.0091 | 36 37 19.2 | −0.059 | 24 | | 14:49–15:17 | |

[a]Total integration time in minutes during the VLA program AR 255 on April 24, 1992.
[b]Total integration time in minutes during the VLA program AR 318 on June 2/3, 1994.
[c]Coordinates of XY Leo as for the VLA detection in Vilhu et al. (1988). This system was not analysed here - see text.
[d]Proper motion determined by Klemola (1977).

Table 2. Results of the VLA Observations

| Obs.# | System | Year | $\alpha_{1950}$ | $\delta_{1950}$ | $f_R$[a] | rms/beam |
|---|---|---|---|---|---|---|
| 1 | W UMa | 1994 | | | | 0.018 |
| 2 | XY Leo[b] | 1994 | | | | 0.016 |
| 3 | AW UMa | 1994 | | | | 0.019 |
| 4 | CC Com | 1994 | | | | 0.014 |
| 5 | AH Vir | 1994 | | | | 0.017 |
| 6 | 44i Boo | 1992 | 15 02 06.61 | +47 50 55.0 | 0.17 | 0.027 |
|   |         |      | 0.05        | 0.4         |      |       |
| 7 | 44i Boo | 1994 | | | | 0.017 |
| 8 | V502 Oph | 1992 | | | | 0.020 |
| 9 | V566 Oph | 1992 | | | | 0.018 |
| 10 | OO Aql | 1992 | 19 45 48.52 | +9 11 01.2 | 0.12 | 0.017 |
|   |        |      | $-0.01$     | $-0.2$     |      |       |
| 11 | OO Aql | 1994 | | | | 0.018 |
| 12 | VW Cep | 1992 | 20 38 06.78 | +75 25 21.5 | 0.35 | 0.022 |
|   |        |      | $-0.01$     | $-0.4$      |      |       |
| 13 | VW Cep | 1994 | 20 38 06.84 | +75 25 22.4 | 0.22 | 0.019 |
|   |        |      | $-0.13$     | $-0.6$      |      |       |
| 14 | SW Lac | 1992 | | | | 0.019 |
| 15 | AB And | 1992 | | | | 0.020 |

[a]Flux density expressed in mJy. To estimate errors and upper limits of the fluxes, use 2× or 4× rms/beam, respectively.

[b]XY Leo was not analysed here – see text.

Notes to Table 2.

Coordinates of the detected systems are for the epoch of observations, in the B1950.0 system. Numbers below the coordinates give shifts of the detected radio sources from the expected, proper-motion corrected positions of the stars.

Table 3. Stellar Parameters and Derived Radio Quantities

| Obs.# | System | $V$ | $M_V$ | $B-V$ | $P$(day) | $\log L_R$ | $\log L_R/L_{bol}$ | Sp |
|---|---|---|---|---|---|---|---|---|
| 1 | W UMa | 7.8 | 4.2 | 0.66 | 0.334 | <14.34 | $<-19.50$ | G |
| 3 | AW UMa | 7.1 | 2.7 | 0.36 | 0.439 | <14.74 | $<-19.67$ | F |
| 4 | CC Com | 11.3 | 7.1 | 1.24 | 0.221 | <14.49 | $<-18.45$ | K |
| 5 | AH Vir | 8.9 | 4.5 | 0.78 | 0.408 | <14.65 | $<-19.09$ | K |
| 6 | 44i Boo | 5.9 | 4.6 | 0.7: | 0.268 | 13.81 | $-19.87$ | K |
| 7 | 44i Boo | | | | | <13.41 | $<-20.26$ | |
| 8 | V502 Oph | 8.3 | 3.9 | 0.66 | 0.453 | <14.75 | $<-19.22$ | G |
| 9 | V566 Oph | 7.5 | 2.9 | 0.39 | 0.410 | <14.77 | $-19.56$ | F |
| 10 | OO Aql | 9.2 | 4.2 | 0.76 | 0.507 | 15.14 | $-18.72$ | K |
| 11 | OO Aql | | | | | <14.92 | $<-18.94$ | |
| 12 | VW Cep | 7.3 | 5.2 | 0.85 | 0.278 | 14.45 | $-19.04$ | K |
| 13 | VW Cep | | | | | 14.24 | $-19.24$ | |
| 14 | SW Lac | 8.5 | 4.6 | 0.75 | 0.321 | <14.50 | $<-19.19$ | K |
| 15 | AB And | 9.5 | 5.2 | 0.9: | 0.332 | <14.71 | $<-18.82$ | G: |

Notes to Table 3.

$L_R$ is in erg s$^{-1}$ Hz$^{-1}$ and $L_R/L_{bol}$ is in Hz$^{-1}$.

TABLE 4. Observations of VW Cep (sorted by radio flux)

| $f_R$ (mJy) | integr (min) | cumul (min) | cumul (frac) | reference |
|---:|---:|---:|---:|---:|
| 8.58 | 5.0 | 5.0 | 0.004 | 3 |
| 4.79 | 9.5 | 14.5 | 0.012 | 1 |
| 4.57 | 9.5 | 24.0 | 0.020 | 1 |
| 4.44 | 5.0 | 29.0 | 0.024 | 3 |
| 3.84 | 9.5 | 38.5 | 0.032 | 1 |
| 3.35 | 5.0 | 43.5 | 0.036 | 3 |
| 3.21 | 9.5 | 53.0 | 0.044 | 1 |
| 2.34 | 5.0 | 58.0 | 0.048 | 3 |
| 1.90 | 5.0 | 63.0 | 0.052 | 3 |
| 1.52 | 30.0 | 93.0 | 0.077 | 3 |
| 1.46 | 30.0 | 123.0 | 0.102 | 3 |
| 1.37 | 9.5 | 132.5 | 0.110 | 1 |
| 1.26 | 60.0 | 192.5 | 0.159 | 3 |
| 0.86 | 8.5 | 201.0 | 0.166 | 1 |
| 0.77 | 8.5 | 209.5 | 0.173 | 1 |
| 0.76 | 9.5 | 219.0 | 0.181 | 1 |
| 0.70 | 21.5 | 240.5 | 0.199 | 1 |
| 0.67 | 7.0 | 247.5 | 0.205 | 1 |
| 0.66 | 5.0 | 252.5 | 0.209 | 3 |
| 0.63 | 9.5 | 262.0 | 0.217 | 1 |
| 0.58 | 60.0 | 322.0 | 0.266 | 3 |
| 0.58 | 21.0 | 343.0 | 0.284 | 1 |
| 0.55 | 9.5 | 352.5 | 0.292 | 1 |
| 0.51 | 8.5 | 361.0 | 0.299 | 1 |
| 0.50 | 22.5 | 383.5 | 0.317 | 1 |
| 0.43 | 60.0 | 443.5 | 0.367 | 3 |
| 0.39 | 22.0 | 465.5 | 0.385 | 1 |
| 0.35 | 60.0 | 525.5 | 0.435 | 3 |
| 0.35 | 24.0 | 549.5 | 0.455 | 4 |
| 0.32 | 60.0 | 609.5 | 0.504 | 3 |
| 0.30 | 60.0 | 669.5 | 0.554 | 3 |
| 0.29 | 8.5 | 678.0 | 0.561 | 1 |
| 0.29 | 60.0 | 738.0 | 0.610 | 3 |
| 0.27 | 30.0 | 768.0 | 0.635 | 3 |
| 0.27 | 60.0 | 828.0 | 0.685 | 3 |
| 0.22 | 31.0 | 859.0 | 0.711 | 4 |
| 0.18 | 60.0 | 919.0 | 0.760 | 3 |
| 0.17 | 60.0 | 979.0 | 0.810 | 3 |
| 0.11 | 230.0 | 1209.0 | 1.000 | 2 |

Notes to Table 4.

References: (1) Hughes & McLean (1984) 6 cm; (2) Rucinski & Seaquist (1988) 6 cm; (3) Vilhu et al. (1988) 6 cm; (4) present, 3.6 cm.

4